\documentclass[fleqn,usenatbib]{mnras}

\usepackage{newtxtext,newtxmath}

\usepackage[T1]{fontenc}

\DeclareRobustCommand{\VAN}[3]{#2}
\let\VANthebibliography\thebibliography
\def\thebibliography{\DeclareRobustCommand{\VAN}[3]{##3}\VANthebibliography}

\usepackage{graphicx}	
\usepackage{amsmath}	
\usepackage{xspace}

\newcommand{\chandra}{\textit{Chandra}\xspace}
\newcommand{\hst}{\textit{HST}\xspace}
\newcommand{\halpha}{H$\alphaup$\xspace}
\newcommand{\iontb}[3]{\ion{#1}{#2}\,\ensuremath{\lambdaup #3}}
\newcommand{\iontbforb}[3]{[\ion{#1}{#2}]\,\ensuremath{\lambdaup #3}}
\newcommand{\cloudy}{\texttt{Cloudy}\xspace}
\newcommand{\pycloudy}{\texttt{pyCloudy}\xspace}
\newcommand{\mappings}{\texttt{Mappings~III}\xspace}
\newcommand{\src}{NGC~6946~X-1\xspace}
\newcommand{\lsrcsoft}{$L_\mathrm{src,soft}$\xspace}
\newcommand{\lsrchard}{$L_\mathrm{src,hard}$\xspace}
\newcommand{\heii}{\iontb{He}{ii}{4686}\xspace}
\newcommand{\hbeta}{\iontb{H}{$\beta$}{4861}\xspace}

\newcommand{\nii}{\iontbforb{N}{ii}{6583}\xspace}
\newcommand{\siione}{\iontbforb{S}{ii}{6716}\xspace}
\newcommand{\siitwo}{\iontbforb{S}{ii}{6731}\xspace}

\title[]{Exploring the case for hard-X-ray beaming in NGC~6946~X-1}

\author[T. Beuchert et al.]
   {Tobias Beuchert$^{1,2}$\thanks{Corresponding E-mail: tobias.beuchert@gv.mpg.de}\thanks{This research has been conducted at the University of Amsterdam},
Matthew J. Middleton$^{3}$,
Roberto Soria$^{4,5,6}$,\newauthor
James C.\,A. Miller-Jones$^{7}$,
Thomas Dauser$^{8}$,\newauthor
Timothy P. Roberts$^{9}$,
Rajath Sathyaprakash$^{10}$,
and Sera Markoff$^{2}$
\\\\
$^{1}$Max Planck Society, Communication Department, Hofgartenstra{\ss}e 8, 80539 M\"unchen, Germany\\
$^{2}$Anton Pannekoek Institute for Astronomy, University of Amsterdam, Science Park 904, 1098 XH Amsterdam, The Netherlands\\
$^{3}$Department of Physics and Astronomy, University of Southampton, Highfield, Southampton SO17 1BJ, UK\\
$^{4}$INAF, Osservatorio Astrofisico di Torino, Strada Osservatorio 20, I-10025 Pino Torinese, Italy\\
$^{5}$Sydney Institute for Astronomy, School of Physics A28, The University of Sydney, Sydney, NSW 2006, Australia\\
$^{6}$College of Astronomy and Space Sciences, University of the Chinese Academy of Sciences, Beijing 100049, China\\
$^{7}$International Centre for Radio Astronomy Research, Curtin University, GPO Box U1987, Perth, WA 6845, Australia\\
$^{8}$Dr. Remeis Sternwarte \& ECAP, University Erlangen-N\"urnberg, Sternwartstrasse 7, 96049 Bamberg, Germany\\
$^{9}$Centre for Extragalactic Astronomy, Durham University, Dept of Physics, South Road, Durham DH1 3LE, UK\\
$^{10}$Istituto Universitario di Studi Superiori, 27100 Pavia, Italy
}

\date{Accepted XXX. Received YYY; in original form ZZZ}

\pubyear{2024}

\begin{document}
\label{firstpage}
\pagerange{\pageref{firstpage}--\pageref{lastpage}}
\maketitle

\begin{abstract}
In order to understand the nature of super-Eddington accretion we must explore both the emission emerging directly from the inflow and its impact on the surroundings. In this paper we test whether we can use the optical line emission of spatially resolved, ionized nebulae around ultraluminous X-ray sources (ULXs) as a proxy for their X-ray luminosity.  We choose the ULX NGC 6946 X-1 and its nebula, MF16, as a test case. By studying how the nebular optical line emission responds to assumed irradiation, we can infer the degree to which we require the UV or X-ray emission from the inflow to be collimated by optically thick winds seemingly ubiquitously associated with ULXs. We find that the nebula is highly sensitive to compact UV emission but mostly insensitive to hard X-rays. Our attempts to quantify the beaming of the soft and hard X-rays therefore strongly depends on the UV luminosity of the ULX in the center of the nebula. We find that it is not possible to conclude a lack of geometrical beaming of hard X-rays from such sources via nebula feedback. 
\end{abstract}




\section{Introduction}

Ultraluminous X-ray sources (ULXs: see the reviews of \citealt{Roberts2007}, \citealt{Kaaret2017}, \citealt{King2023}), are known to contain stellar remnants in the form of neutron stars and probably stellar mass black holes (although the relative number remains unknown, see e.g. \citealt{Middleton2016, King_Lasota2020}). In the simplest situation where the accreting compact object is a black hole or a low dipole field neutron star, the intrinsic luminosity can be increased at most by a factor of a few relative to the Eddington limit \citep{Shakura1973}. Mechanisms are therefore needed to explain observed luminosities of $>$10$^{40}$\,erg\,s$^{-1}$. One of the simplest and most direct methods has been geometrical collimation (beaming) of the innermost regions \citep{King2009}. Such collimation is a natural corollary of optically thick winds accompanying accretion above tens of times the Eddington limit \citep{Shakura1973,Poutanen2007}, seen in global MHD simulations \citep{Ohsuga2011}, and detected in X-ray observations \citep{Middleton2014,Middleton2015a,Middleton2015b,Walton2016,Pinto2016,Pinto2017,Kosec2018,Kosec2021}.

It is expected that the innermost regions of ULXs are the most collimated and so naturally the hardest X-ray emission should tend to be the most beamed (which is supported by observations of ULXs tending to become harder when brighter: \citealt{Weng2018}). The picture can become somewhat more complicated when the compact object is a neutron star, especially where the magnetic field strength is high (typically $>$ 10$^{13}$\,G) or accretion rate low enough such that the accretion flow never reaches the local Eddington limit in the disc \citep[e.g.,][]{Mushtukov2015, Mushtukov2017}. Mass loss is still likely in even these cases, as the fan-beam is expected to produce super-Eddington luminosities which impinge on material in the magnetosphere (e.g. \citealt{Mushtukov2024}). Assuming beaming acts to some degree and is accretion rate dependent, the anisotropy and amplification of flux must affect our ability to observe the population of ULXs \citep{Middleton2016,Wiktorowicz2019} which will also be further influenced by precession of the disc/wind \citep{Middleton2015b,Middleton2018,Middleton2019,Khan2022}.

There have been many attempts to quantify the collimation/beaming of ULXs. One of the primary tools has been the use of surrounding nebulae of ionised gas (the ULX bubble nebulae, e.g. \citealt{Pakull2002}) as calorimeters. The lines from such nebulae are excited by a combination of shocks by winds and jets, and photo-ionisation by the bright central source (e.g. \citealt{Gurpide2022}). Photo-ionisation modelling allows the spectral energy distribution (SED) seen by the nebula to be compared to the observed SED of the central ULX; the ratio of the two then implies the beaming factor. Such an approach led \citet{Pakull2002} to determine that, in the case of the well-studied ULX, Ho II X-1, the luminosity seen by the nebula was similar to the X-ray luminosity we observe (a similar finding was noted in the case of NGC 5408 X-1: \citealt{Kaaret2009}). Ho II X-1 is one of the softest ULXs \citep[see, e.g., ][]{Middleton2015a} and may be seen at higher inclinations \citep{Poutanen2007}. As the wind preferentially collimates the hardest X-rays, this should not be seen to imply there is no beaming in other ULXs which may be considerably spectrally harder \citep{Gladstone2009,Middleton2015a,Pintore2017,Koliopanos2017,Walton2018}. 

In this paper we model the emission from the photo-ionised nebula MF16 \citep{Matonick1997,Kaaret2010,Long2019} around the ULX \src\ \citep{Roberts2003}. By performing a detailed photo-ionisation modelling, we test whether the nebula can in fact constrain the geometrical beaming of this source (as well as similar ULXs).

\section{\src\ and the MF16 nebula}

\src\ is one of the spectrally softest ULXs (very similar in nature to NGC 5408 X-1) with an observed (absorbed) X-ray luminosity ranging between 3-4$\times$10$^{39}$ erg/s \citep{Middleton2015a} at a distance of 7.7\,Mpc \citep{Anand2018,Murphy2018,Eldridge2019}. It is notable for being one of the most X-ray variable ULXs, with fractional rms values within single observations reaching 40\,\% \citep{Middleton2015a} and QPOs reported at 10s of mHz \citep{Rao2010}. The surrounding MF 16 nebula is thought to be both photo-ionised and shock heated (\citealt{Abolmasov2008}). Intriguingly, UV observations of this source by the {\it Hubble Space Telescope} (\hst) imply the source to be an ultra-luminous UV source \citep{Kaaret2010} with this radiation able to explain the the observed \heii\ emission-lines \citep{Abolmasov2008}. MF16 is also one of only a small number of nebulae detected in the IR by Spitzer, suggesting the presence of a red super-giant donor star \citep{Berghea2012}.

\section{Observations and Spectral Analysis}
\label{sec:obsspec} 

In this section we describe NGC 6946 X-1 using photometric UV and spectroscopic X-ray data, and the response of the surrounding MF16 nebula using radio interferometry and optical photometry.

\subsection{Imaging the MF16 nebula}
\begin{figure}
  \includegraphics[width=\columnwidth]{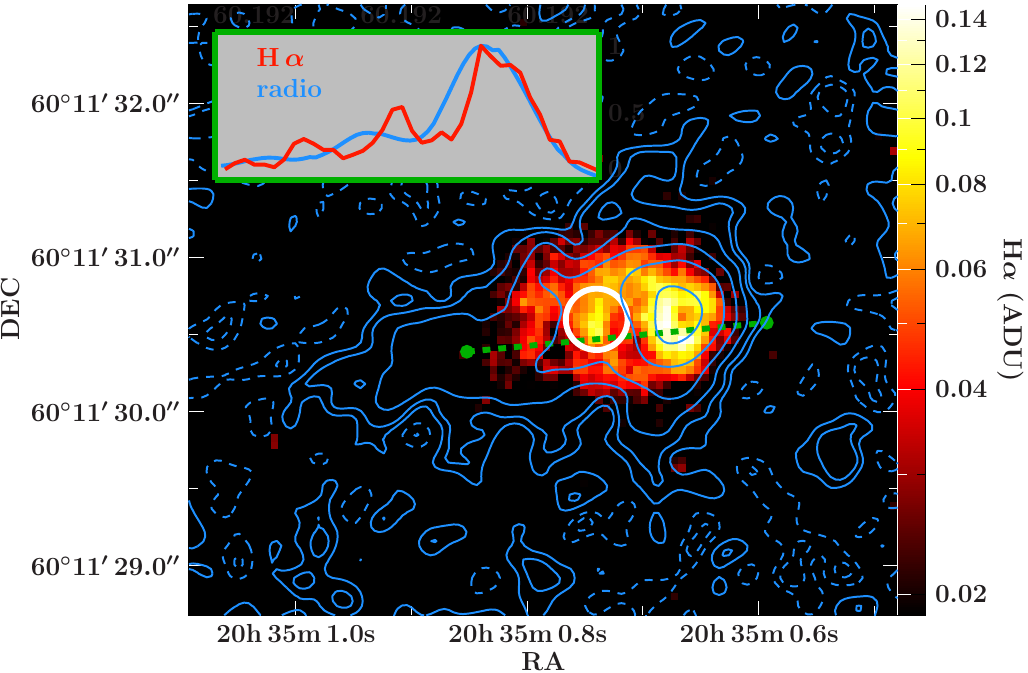}
  \caption{\hst\ map of MF16 in \halpha\ (F656N filtered emission in color scale) with overlaid, stacked C- and X-band VLA contours. The radio contour levels (blue) are (-8.46, -4.23, -2.11, 2.11, 4.23, 8.46, 16.92, 33.85, 67.69)\,$\upmu$Jy/beam starting at the rms noise level of 2.11\,$\upmu$Jy/beam. The synthesized beam of the stacked image measures 0.25"$\times$ 0.30" The white circle is centred on the position of the X-ray  source, with the radius corresponding to the 0.2" positional uncertainty of the averaged \chandra\ pointings used in this work. The inset shows intensity profiles for the \halpha\ and radio maps along the green dashed line. Both profiles are normalized to a common maximum.}
  \label{fig:map}
\end{figure}
The superb spatial resolution of \chandra, \hst, and the NRAO's Karl G. Jansky Very Large Array (VLA) allow us to trace the compact object's feedback independently via direct imaging. In Fig.~\ref{fig:map} we overlay maps of \halpha\ and a stack consisting of 5\,GHz and 10\,GHz radio maps. 

\subsubsection{Imaging with \textit{HST}}
We aligned the \halpha\ image (\hst/WFPC2/F656N taken in 1996 January 27) to the absolute astrometric frame of reference. Due to the low number of catalogued stars in this narrow-field image (less than three), we first matched it to another, wider-field \hst/ACS/WFC image. We astrometrically corrected this image using 81 reference stars from the URAT1 catalogue \citep{Zacharias2015}. The astrometric solution is accurate to within 0\farcs1. 

The \halpha\ nebula has a shape very similar to that seen in [\ion{S}{ii}] \citep{Blair2001}, both of which are usually enhanced under shocked conditions \citep[see][and Fig.~\ref{fig:results} for \ion{[Si}{ii]}]{Allen2008}. The overall structure measures $1\farcs4\times 0\farcs8$, i.e., $50\times 30$\,pc at a distance of 7.7\,Mpc \citep{Anand2018,Eldridge2019,Murphy2018}\footnote{This value was derived from measurements of the tip of the red-giant branch which we consider to be more accurate than previous measurements of around 5--6\,Mpc used, e.g., by \cite{Abolmasov2008}}.

\subsubsection{Radio imaging with the VLA}
The radio data originate from an observing campaign with the VLA under program code 15A-133.  The first two observing epochs (2015 June 20, 2015 June 30) were taken at 8--12\,GHz, and the second two epochs (2015 July 10, 2015 July 18) at 4--8\,GHz, with 4\,GHz of bandwidth in each frequency band. The array was in its most-extended A-configuration for all observations.
We used either 3C\,48 or 3C\,286 to calibrate the delays and the instrumental frequency response, and to set the amplitude scale. For the complex gain calibration, we used the nearby calibrator J2022+6136 ($2\fdg 1$ from MF16). Data were processed according to standard procedures using the Common Astronomy Software Application \citep[CASA v4.4.0,][]{CASA2022}. To maximise the image sensitivity, we imaged all four epochs together, using two Taylor terms to account for the frequency-dependence of the sky brightness.

While the orientation of the radio nebula is consistent with the \halpha\ map, the radio emission to the east has no \halpha\ counterpart. The \halpha\ nebula is brightest towards the west, and coincides with the peak of the radio-intensity map. The intensity profiles of both \halpha\ and radio emission along the green dashed line in Fig.~\ref{fig:map} trace each other well, down to a radio brightness level of 8.46\,$\upmu$Jy/beam, which might imply that shocks are a common feedback process at these scales.

\subsubsection{Pinpointing the X-ray emission with \textit{Chandra}}
In Fig.~\ref{fig:map}, we indicate the X-ray emitting region and its positional uncertainty of 0.2". This region coincides with the secondary peak of the \halpha\ nebula and is, within uncertainties, consistent with the geometrical center of the \halpha\ nebula. We determine this co-location from relative astrometry, using the point-like X-ray source SN2017eaw and its optical counterpart, surrounded by numerous reference stars seen with \hst\ (F606W and F814W images from 2016, and the WFC3 images from 2018). 
The \chandra\ position of \src\ lies within $\sim 4-5$\arcmin\ of the aimpoint for all observations except for \texttt{obsid} 19040, at an offset of 2\farcm3 and for \texttt{obsid} 4404 with an offset of 2\farcm9. 

In every \chandra/ACIS observation, the fitted extent of the X-ray source is consistent with the point spread function (PSF) of the ACIS detector at that off-axis location. Therefore, we cannot spatially distinguish between the point-like emission from the compact object (ULX) and thermal plasma emission from ionized gas around it (either in a fast outflow or in the nebula).

\subsection{Spectral observations and data reduction}
\label{sect:obs} 
In order to quantify and isolate the UV-to-X-ray spectrum from the compact ULX, we make use of the unprecedented sub-arcsec resolution of \hst\ and \chandra.

\subsubsection{X-ray spectroscopy with \textit{Chandra}}
NGC\,6946 was observed by \chandra/ACIS on nine occasions between September 2001 and June 2017, with a total exposure time of $\approx$261 ks (Table~\ref{tab:obs}). After downloading the data from the public archive, we reprocessed and analysed them with the Chandra Interactive Analysis of Observations ({\sc ciao}) software version 4.13 \citep{Fruscione2006}. We created new level-2 event files with the {\sc ciao} task {\it chandra\_repro}, generated stacked images in different energy bands with {\it merge\_obs} and {\it dmcopy}. We used {\it srcflux} to determine model-independent fluxes from the event files, and {\it specextract} to extract spectra and associated response and ancillary response files for individual observations.  In principle, the emitting region is likely to be spatially extended, with X-ray contributions from the compact object and the surrounding hot gas in the MF16 nebula. However, as discussed, the extent of the X-ray emission is always consistent with the ACIS PSF (Section 3.1.3); we therefore selected {\it specextract} parameters suitable for point-source extraction (`correctpsf = yes'). 
\begin{table}
  \centering
  \caption{List of \chandra\ observations used in this work.}
  \label{tab:obs}
  \begin{tabular}{llll} 
    \hline
    Obsid & date & exposure\,[ks] & counts [$\times 10^{3}$]\\
    \hline
    1043 & 2001-09-07 & 58.2 & 8.7 \\
    4404 & 2002-11-25  & 30.0 & 3.9 \\
    4631 & 2004-10-22  & 29.7  & 3.2 \\
    4632 & 2004-11-06  & 28.0  & 3.7 \\
    4633 & 2004-12-03  & 26.6  & 3.4 \\
    13435 & 2012-05-21  & 20.4  & 1.8 \\
    17878 & 2016-09-28  & 40.0  & 2.8 \\
    19887 & 2016-09-28   & 18.6  & 1.3 \\    
    19040 & 2017-06-11  & 10.0  & 0.7 \\
    \hline
  \end{tabular}
\end{table}

\subsubsection{UV photometry with \textit{HST}}
For the UV part of the spectrum, we adopt \hst\ point-source fluxes from May 2008 as listed in \citet{Kaaret2010} for the Wide Field and Planetary Camera 2, WFPC2, with the filters F450W, F555W and F814W as well as the ACS Wide Field Camera, WFC, with the filter F140LP. We use the bandwidths\footnote{\url{https://www.stsci.edu/hst/documentation/handbook-archive}} 127\,\AA, 951\,\AA, 1228\,\AA\ and 1539\,\AA, respectively, to further process these data. We note that the UV and time-averaged X-ray data used in this work are non-simultaneous. The X-ray spectrum is observed to vary (mostly above 1~keV) on long timescales but only between a 0.3--10 keV deabsorbed luminosity of $\sim$ 0.9-1.3$\times$10$^{40}$ erg/s assuming an intrinsic column of $2.90\times 10^{21}\,\mathrm{cm}^{-2}$ (Sect.\,\ref{sect:spectral}) at 7.7\,Mpc \citep{Middleton2015a}. Given the relatively small dynamic range in the spectrum (accepting that this is could be an underestimate due to incomplete sampling of the source behaviour), we consider it acceptable to average over these variations for our feedback study.

In order to be able to fit the broad-band spectrum including the UV and X-ray data with both extinction and photoabsorption applied, we re-redden the UV data with $A_\mathrm{V}=1.54$ \citep{Abolmasov2008}. While \citet{Kaaret2010} used the extinction curve by \citet{Cardelli1989} for de-reddening, we implemented the parameterized curve by \citet{Fitzpatrick1999} in \texttt{ISIS} \citep{Houck2000}. The difference between both curves is negligible for our purpose \citep[see Fig.~1 in][]{Fitzpatrick1999}. 

\subsubsection{Spectral lines with the Subaru SAO 6-m telescope}
We make use of six characteristic line transitions measured with the SAO 6-m telescope \citep{Abolmasov2008} in order to assess the nebular response to the ULX SED, i.e., \iontb{He}{ii}{4686}, \iontb{H}{$\beta$}{4861}, \iontbforb{O}{iii}{5007}, \iontbforb{N}{ii}{6583}, \iontbforb{S}{ii}{6716}, and \iontbforb{S}{ii}{6731}. The intensities of the first three lines were measured with the SCORPIO long-slit spectrograph \citep{Afanasiev2005} in 2005 June 10, and the latter three lines with the MultiPupil Fiber Spectrograph (MPFS: \citealt{Afanasiev2001}) in 2005 January 17 \citep{Abolmasov2008}. The de-reddened intensities (assuming $A_\mathrm{V}=1.54$) are provided in Table~\ref{tab:lines}. We also present a value for \iontb{H}{$\alpha$}{6563} as we know the \ion{H}{$\alpha$}/\ion{H}{$\beta$} flux ratio to be $\sim$2.8 from \cite{Abolmasov2008} (and propagate errors accordingly).

We consider a contribution to the emitted line intensities by shocked gas.
In a previous study, \citet{Berghea2012} directly assessed the influence of shocks using infrared lines around \ion{O}{iv}, but also using \heii. Photoionized gas, modeled with \cloudy can reproduce the observed line intensities well, and only shocks much faster than $300\,$km/s would provide enough ionizing potential to yield a similar match. As the authors note, the shock speed of $\sim 250\,$km/s estimated by \citet{Dunne2000} is not enough to justify shock ionization as the main feedback channel. To further confirm this, we use \mappings\ (noting that the \mappings\ code assumes a steady-state shock) and tabulated line ratios relative to \hbeta\ \citep{Allen2008} to predict line intensities for a shock speed of 250\,km/s and find that, even if such fast shocks could reproduce \heii, both \nii\ and \ion{S}{ii} would be largely overproduced. We therefore concur with \citet{Berghea2012} that the observed \heii\  line intensity can not be reproduced solely by shock-ionization, a combination of photoionization and shock-ionization, however, remains very plausible. 

For the shocked component, we use the \texttt{Mappings III} library \citep{Allen2008} and consider the emission from both the shock moving at 250\,km/s and its precursor for weakly magnetized gas ($B/n^{2}=0.1\,\mu\mathrm{G}\,\mathrm{cm}^{3/2}$), which is a valid assumption for shocked gas at considerable distance from the compact source. The intensities of the lines in question, emerging from the shock, are found to be unchanged below the chosen magnetization parameter $B/n^{2}$. We list the theoretically assumed intrinsic line intensities for shocked gas in \src in Table~\ref{tab:lines}.
\begin{table}
  \centering
  \caption{De-reddened, intrinsic line intensities \citep{Abolmasov2008} and the theoretically assumed shocked portion in units of $\times 10^{-16}\,\mathrm{erg}\,\mathrm{cm}^{-2}\,\mathrm{s}^{-1}$. We apply $A_\mathrm{V}=1.54$ for de-reddening.}
  \label{tab:lines}
  \begin{tabular}{lll} 
    \hline
    Line & $F_\mathrm{deredd}$ & $F_\mathrm{shocked}$ \\
    \hline
    \iontb{He}{ii}{4686} & $56\pm 2$ & 9 \\
    \iontb{H}{$\beta$}{4861} & $249\pm4$ & 159 \\
    \iontbforb{O}{iii}{5007} & $1620\pm 30$ & 242 \\
    \iontb{H}{$\alpha$}{6563} & $687\pm17$ & \\
    \iontbforb{N}{ii}{6583} & $611\pm12$ & 194 \\
    \iontbforb{S}{ii}{6716} & $352\pm3$ & 103 \\
    \iontbforb{S}{ii}{6731} & $335\pm4$ & 172  \\
    \hline
  \end{tabular}
\end{table}

\subsection{Spectral Analysis}

\label{sect:spectral}
\begin{figure}
  \includegraphics[width=\columnwidth]{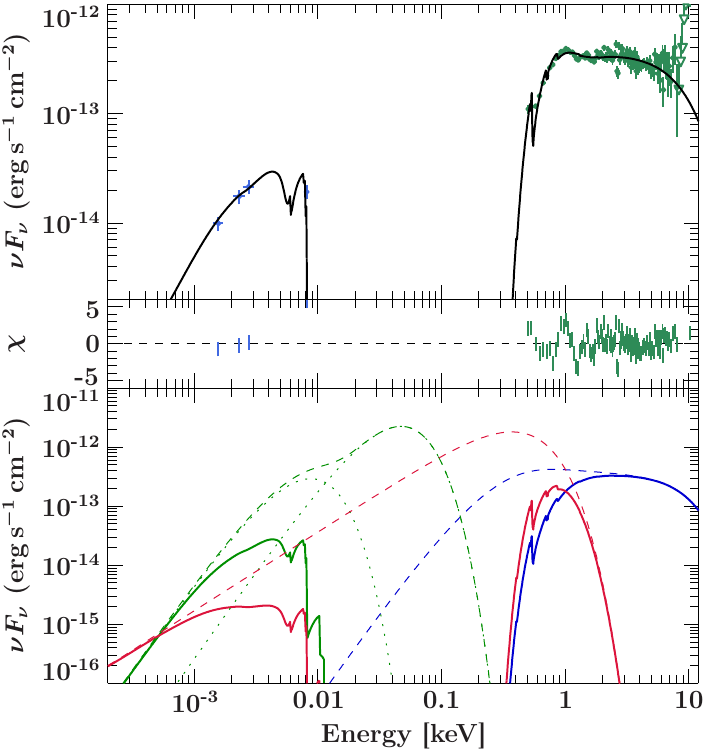}
  \caption{Broad-band UV-to-X-ray spectrum for the compact
emission of \src. \textit{Top panel}: \hst\ photometric data
(blue) and the {\it Chandra} spectrum (green) in $\nu F\nu$ units.
The black line corresponds to the best-fit spectral model as described in the text. The residuals to this fit are shown in the panel
below. Clear residuals are detected which correspond to broadened line features imprinted by outflows (Middleton et al. 2014, 2015; Pinto et al. 2016). \textit{Bottom panel}: Individual absorbed / reddened (solid lines) and de-absorbed / de-reddened model components (dashed lines): \texttt{bbody} (green), \texttt{diskbb} (red), \texttt{nthcomp} (blue).}
  \label{fig:sed}
\end{figure}

\begin{figure}
  \includegraphics[width=\columnwidth]{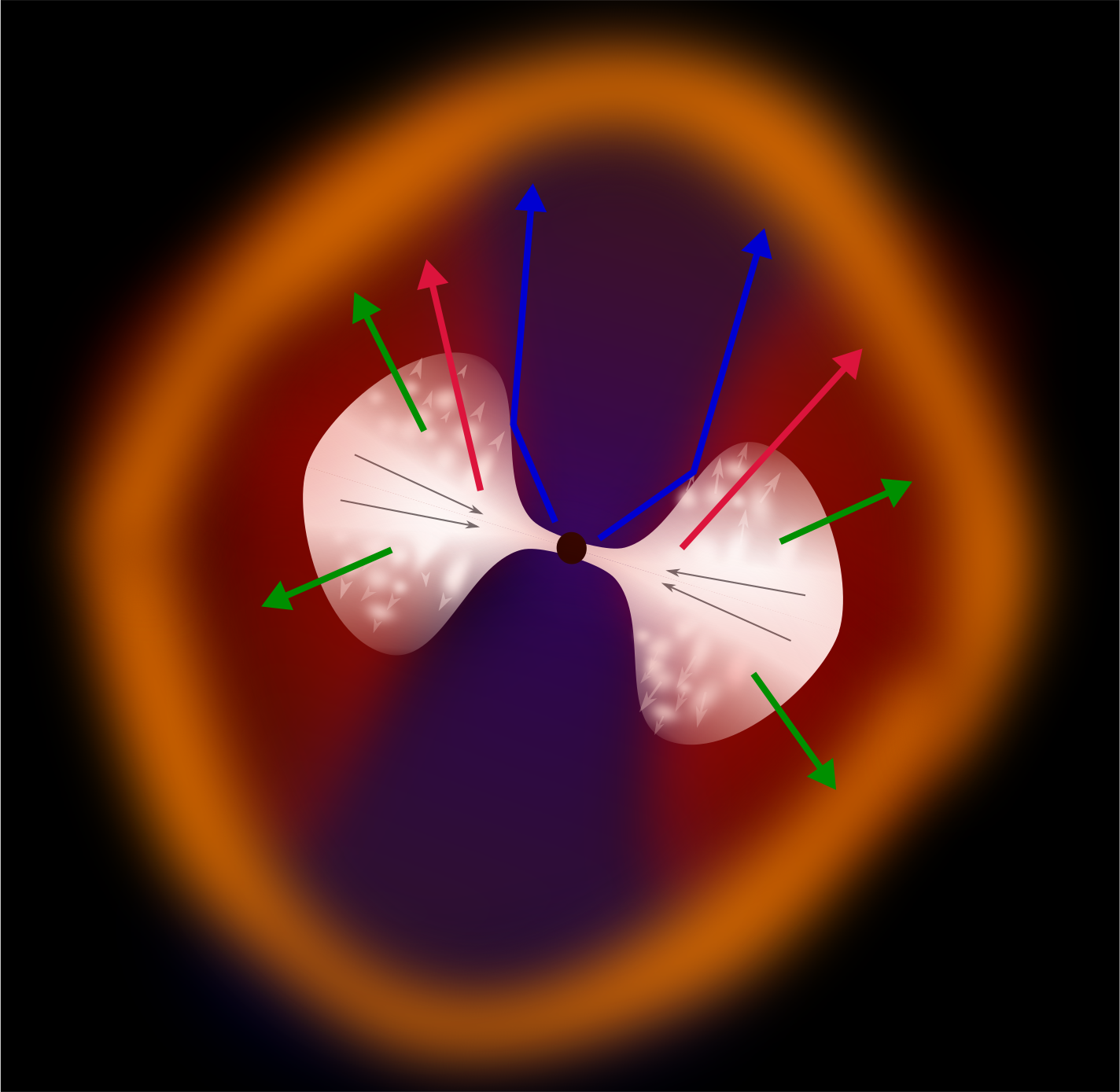}
  \caption{Illustration of the accreting environment of \src, which is the source of the radiation illuminating the surrounding MF16 nebula. The arrows denote the radiation emitted by the outer face of an extended photosphere (green), that generated within the flow and emerging at smaller inclinations (red) and those from the innermost regions which are most likely to be reflected from the cone defined by the wind (blue). The spectral components of Fig.~\ref{fig:sed} follow the same colour code.}
  \label{fig:modelnebula}
\end{figure}

We show the broad-band spectrum of the ULX in Fig.~\ref{fig:sed}, which is assumed to illuminate the surrounding nebula. We base our modeling on recent advances in the phenomenological understanding of individual spectral components and their timing behavior in the context of a geometrically thick accretion flow that is dominated by optically thick winds \citep{Middleton2015a}. The assumed geometry of the nebula is illustrated in Fig.~\ref{fig:modelnebula}. We distinguish three separate emission components that originate from the outermost photosphere (UV continuum), the photosphere of the disk-wind at smaller radii/inclinations (soft X-ray spectral component), and a hard spectral component coming from the innermost regions that is most collimated by the disk wind. According to the assumed geometry, these components not only cover distinct parts of the spectrum but also illuminate distinct parts of the nebula.

We model the \hst\ data with a dual blackbody spectrum (green model components in Fig.~\ref{fig:sed}) with $T_{1}=0.012$\,keV ($10^{5.15}\,$K) and $T_{2}=0.00224$\,keV ($10^{4.41}$\,K), as suggested by \citet{Kaaret2010}. \citet{Kaaret2010} explain the cooler UV blackbody as emission from an O or B-type companion star (although it may also originate from the outer photosphere of the wind: \citealt{Poutanen2007}) whilst the hotter blackbody -- characteristic of the extreme UV-bright source \src\ -- may emerge from the outer photosphere or result from X-rays from the inner regions being down-scattered in the wind. We fix the temperature of the cooler blackbody (with $T=T_{2}$) as well as both the temperature and normalization of the hotter blackbody (with $T=T_{1}$) to the literature values from \citet{Kaaret2010} in order to minimize degeneracies while fitting the broad-band UV-to-X-ray spectrum. 

We model the X-ray data using a combination of a disc blackbody ({\sc diskbb} -- red lines in Fig.~\ref{fig:sed}), and a broad, hotter component  ({\sc nthcomp} -- blue lines in Fig.~\ref{fig:sed}). Assuming the ULX to be viewed at moderate inclinations (\citealt{Middleton2015a}), the former component can be associated with soft-X-ray emission from large radii, and the latter with hard-X-ray emission from the innermost, hottest regions of the accretion flow (\citealt{Poutanen2007}). That emission would, in the model of \citealt{Middleton2015a}, be partly downscattered through the wind, and re-emerge as a contribution to the soft continuum. A very similar hot continuum is also needed in the feedback study of \citet{Abolmasov2008} in order to explain the optical line intensities from Table~\ref{tab:lines}).

The total intrinsic continuum model we apply to the UV-X-ray data is given by: \texttt{zbbody(1)+zbbody(2)+diskbb+nthcomp}. We model intrinsic photo-absorption by neutral gas with \texttt{tbnew} with corresponding abundances \citep{Wilms2000}, which is valid down to the EUV. We also account for reddening (\texttt{redden} in \texttt{ISIS}) using the extinction law as parameterized by \citet{Fitzpatrick1999} and the Galactic gas-to-dust ratio $N_\text{H}=A_\text{V}\times 1.79 \times 10^{21}\,\mathrm{cm}^{-2}\,\mathrm{mag}^{-1}$ \citep{Nowak2012} with $A_\mathrm{V}$ frozen to 1.54 \citep{Abolmasov2008}. This translated to a reddening column density of $2.75\times 10^{21}\,\mathrm{cm}^{-2}$. The total photoabsorption column density we obtain from fitting, i.e., the sum of $N_\mathrm{H,Gal}\sim 2.11\times 10^{21}\,\mathrm{cm}^{-2}$ and $N_\mathrm{H,intr}\sim 2.90\times 10^{21}\,\mathrm{cm}^{-2}$, exceeds the reddening column density. This may either imply that the Galactic gas-to-dust ratio used here or the extinction value $A_\text{V}$ are too low. The latter is more plausible given the sparse sampling of the UV continuum. We also note that we are fitting a non-simultaneous SED, whilst the neutral column density may change with time \citep{Middleton2015b}. All best-fit parameters are listed in Table~\ref{tab:specpar}. We obtain a fit statistic of $\chi^2/\mathrm{dof} = 185/61$ which is poor due to line-like residuals below $\sim 2$\,keV (Fig.~\ref{fig:sed}), which are generally well shown to be related to mass-loaded winds in ULXs \citep{Middleton2014,Middleton2015b,Pinto2016,Walton2016,Pinto2017,Kosec2018, Kosec2021}. For the purpose of our work, we only focus on the continuum and do not explore modelling of the line-like features as these do not affect the overall shape of the continuum (see \citealt{Middleton2014}). 

\begin{table}
  \centering
  \caption{Spectral parameters of the best fitting model to the broad-band spectrum of \src. We apply a Galactic absorption column of $N_\mathrm{H,Gal}=2.11\times 10^{21}\,\mathrm{cm}^{-2}$ and a redshift of $z=0.00134$, which, unlike the cosmological redshift, represents the local kinematics of the Virgo cluster, Global Attractor, and Shapley supercluster \citep{Mould2000}. Frozen parameters are denoted with $\dagger$.}
  \label{tab:specpar}
  \begin{tabular}{lll} 
    \hline
    model & parameter & value \\
    \hline
    \texttt{zbbody} & norm & $3.65\times 10^{-5\,\dagger}$ \\
                    & k$T$\,(keV) & 0.012$^\dagger$ \\
                    & norm & $\left(4.7\pm1.0\right)\times10^{-6}$ \\
                    & k$T$\,(keV) & 0.00224$^\dagger$ \\
    \texttt{diskbb} & norm & $\left(2.7^{+2.8}_{-1.5}\right)\times10^{2}$ \\
                    & $T_\mathrm{in}$\,(keV) & $0.157^{+0.016}_{-0.011}$ \\
    \texttt{nthcomp} & norm & $\left(2.59^{+0.25}_{-0.28}\right)\times10^{-4}$ \\
                      & $\Gamma$ & $2.13^{+0.10}_{-0.11}$ \\
                      & k$T_\mathrm{e}$\,(keV) & $2.1^{+1.5}_{-0.5}$ \\
    \texttt{tbnew} & $N_\mathrm{H,intr}$\,($10^{22}\,\mathrm{cm}^{-2}$) & $0.29\pm0.06$ \\
    \texttt{redden} & $N_\mathrm{H,intr}$\,($10^{21}\,\mathrm{cm}^{-2}$) & $2.75^\dagger$ \\
                    & $R$ & 3.1$^\dagger$\\
    \hline
  \end{tabular}
\end{table}

After investigating the decomposition of the {\it observed} ULX continuum, we can exploit the capabilities of the surrounding nebula as a calorimeter to reliably probe the {\it intrinsic} continuum, to which we can make comparisons and test cases for geometrical beaming. 

\subsection{Feedback into the nebula}
\label{sect:feedback}
The \src/MF16 system is well-suited for a study of energy being deposited into the environment. Radiative feedback appears to be the dominant process in this source \citep[e.g.,][]{Berghea2012}, which is supported by the exceptionally strong UV continuum probed by \hst. We expand on previous feedback studies \citep[e.g.,][]{Abolmasov2008} by exploring the feedback separately for the quasi-thermal UV and X-ray components shown in Fig.~\ref{fig:sed}. 

Our immediate goal is to quantify whether the observed luminosity integrated over each component, corresponds to the intrinsic luminosity needed to reproduce the optical lines emitted by MF16. A mismatch between the observed and intrinsic luminosities would likely argue against isotropic irradiation of the nebula and instead favor some level of geometrical beaming.

We note that, in our analysis, we have used a revised distance for \src of 7.7\,Mpc, which is higher than those distances adopted in previous feedback studies. This also results in intrinsic luminosities that are higher than previously assumed. 

Our use of a non-simultaneous UV--X-ray SED containing X-ray data time-averaged over 16\,years can be justified by three main arguments. Firstly, the light travel time to the nebula at $\sim 10\,\mathrm{pc}$ is $\gtrsim 30\,$yr, much longer than the time-scales covered by the gap between individual X-ray observations (Table~\ref{tab:obs}), and the X-ray SED is relatively stable on long timescales \citet{Middleton2015a}. The recombination time scale of the surrounding ionized gas, i.e., $t_\mathrm{rec} \sim 2\times 10^{4}\,Z^{-2}\,T_{5}^{1/2}\,n_{9}^{-1}\,$s \citep{Krolik1999}, is an important measure for the ability of the surrounding gas to react to source-intrinsic variability. We estimate the recombination time using $n_{9}=5.7\times 10^{-7}$ for the electron number density in units of $10^{9}\,\mathrm{cm}^{-3}$, \citep{Abolmasov2008} (the authors derived the electron number density for [SII], which we shall use as a rough guess for the local densities), the temperature derived for [OIII] ($T_{5}=0.18$ in units of $10^{5}\,$K) and the corresponding atomic number of oxygen ($Z=8$). The result of $t_\mathrm{rec}\sim 7\,$yr is on the order of the total baseline for the X-ray observations
and the light-crossing time of the nebula. Finally, should it be the case that the ULX wind is precessing, the associated variability timescales tend to be far shorter than those of the nebula such that the nebula will only respond to the averaged irradiation as we have assumed.

We proceed to use the plasma code \cloudy\  \citep{Ferland2017} to predict the optical line response from our photoionized model nebula, and then make comparisons to the {\it observed} line intensities of MF16. 

As optical spectroscopy probes the entire gas distribution around \src, we can approximate MF16 by a spherical geometry with a filling factor of 0.2 in \cloudy\ \citep{Dunne2000,Blair2001,Berghea2012}. 
We illuminate this nebula with the soft and hard X-ray spectral components as well as with the UV continuum (combining both blackbodies) from Fig.~\ref{fig:sed}, and explore how the nebula response changes with the input luminosities of the separate spectral components. If one spectral component is geometrically beamed, we expect a mismatch of its intrinsic luminosity (denoted as $L_\mathrm{src}$), constrained from our feedback study, and the observed one (denoted as $L_\mathrm{obs}$). In practice, we use \pycloudy\ \citep{Morisset2013} to run \cloudy\ for a range of input luminosities $L_\mathrm{src}$ around the observed values $L_\mathrm{obs}$, i.e., $10^{-2}\,L_\mathrm{obs} \rightarrow 10^{2}\,L_\mathrm{obs}$, while $L_\mathrm{obs}$ is the observed (deabsorbed) luminosity of the soft and hard X-ray components, \texttt{diskbb} ($\log L_\mathrm{obs,soft}=40.41$) and \texttt{nthcomp} ($\log L_\mathrm{obs,hard}=39.98$) from Fig.~\ref{fig:sed}. We repeat this exercise for five different intrinsic UV luminosities, i.e., $\log L_\mathrm{UV} = 39.4, 39.8, 40.4, 40.8, \mathrm{and}~41.4$, to account for the uncertainty concerning the shape and integrated luminosity of the true intrinsic UV continuum. Our best-guess ($\log L_\mathrm{UV} = 40.4$) originates from de-reddening a double-black-body continuum fitted to the sparsely covered UV spectrum used in this work, with a value of $A_\mathrm{V}$ that is uncertain itself \citep{Abolmasov2008}. These luminosities have been derived after integrating the spectrum of the corresponding spectral component between $10^{-5}$\,keV and $10^{5}$\,keV. $L_\mathrm{UV}$ corresponds to the double-black-body spectrum fitted in Fig.~\ref{fig:sed}. As a source of additional isotropic radiation, we include the CMB within the \texttt{Cloudy} model, with input parameters being the electron gas density $n_\mathrm{e}=570\,\mathrm{cm}^{-3}$, derived from the \iontbforb{S}{ii} 6716/6731 line ratio \citep{Abolmasov2008}, and the source distance of $\sim 7.7\,$Mpc. The following assumptions allow us to speed up the code without loss of relevant physics: we neglect molecules for the calculations of the photoionized plasma, and use a coarse opacity grid. We also assume a less comprehensive database of H and He-like ions. For H and He-like ions of iron, we set the number of resolved and collapsed levels to 10 and 15, respectively.  We also neglect elements with relative abundances of $\log [n(\mathrm{X})/n(\mathrm{H})]<-7$.

\begin{figure*}
  \includegraphics[width=\textwidth]{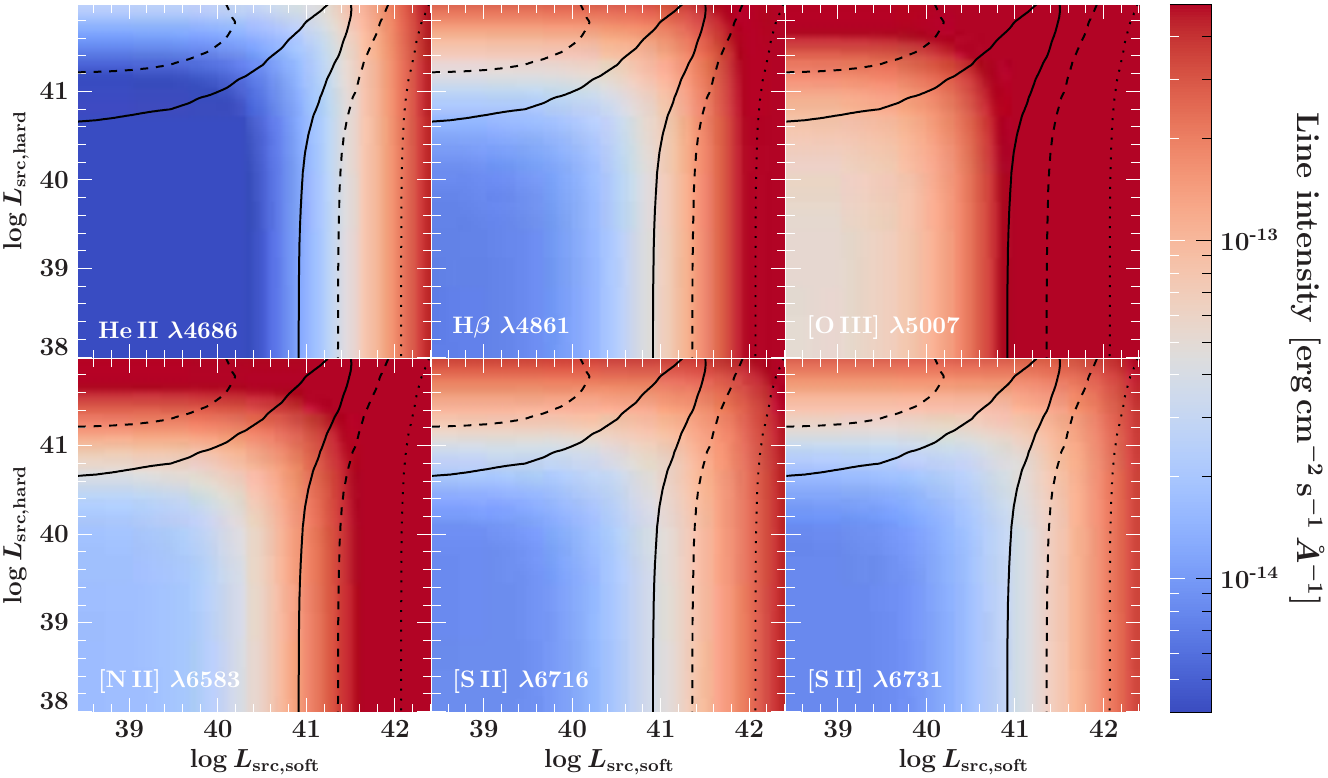}
  \caption{Parameter space of our \cloudy-based table model in \texttt{ISIS}, showing the line intensities of interest as functions of \lsrcsoft\ and \lsrchard for an intrinsic UV continuum with $\log L_\mathrm{UV} = 39.9$. We plot contours for the 68.3\%, 90\%, and 99\% confidence limits (dotted, dashed, and solid lines, respectively) which represent the allowed parameter space for a fit of this tabulated model to the observed line intensities, after adding 30\% systematics and assuming an intrinsic UV luminosity of $\mathrm{log}\,L_\mathrm{UV} = 39.9$.}
  \label{fig:parspace}
\end{figure*}

We integrate the results of the \pycloudy\ runs to make a set of five \texttt{ISIS} table models, one per UV continuum. Each table model returns the expected intensities of the lines from Table~\ref{tab:lines}, with the model parameters \lsrcsoft\ and \lsrchard, i.e., the intrinsic (i.e. not necessarily the observed), de-absorbed luminosity of the spectral components \texttt{diskbb} and \texttt{nthcomp}. Figure~\ref{fig:parspace} illustrates the parameter space in colour scale. The line intensities generally increase with increasing X-ray luminosity, but only become a strong function of these for $\log L_\mathrm{src} \gtrsim 41$. 

\begin{figure*}
  \includegraphics[width=\textwidth]{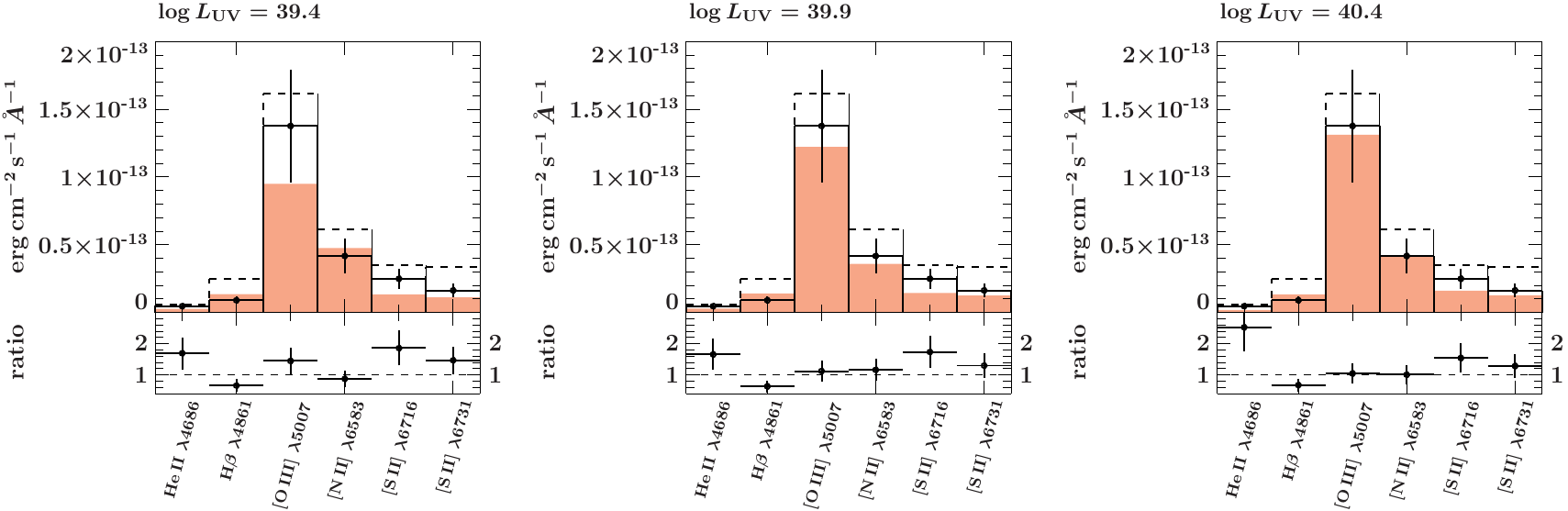}
  \caption{Fit results for our \cloudy-constructed table model (colored bars) together with the observed de-reddened and shock-subtracted line intensities (black solid lines with corresponding data points and uncertainties). The dashed lines correspond to the total line intensities, including a likely shock contribution from Table~\ref{tab:lines}. In the cases shown here, we add an additional} 30\% systematic uncertainty to the statistical uncertainty from Table~\ref{tab:lines}. The bottom panels show data/model ratios. The three panels illustrate fits with different incident UV luminosities, i.e., $\mathrm{log}\,L_\mathrm{UV}=39.4, 39.9$, and 40.4, respectively. These fits yield  $\chi^{2}\,(\mathrm{dof}) = 9.5\,(3), 8.2 \,(3), \mathrm{and}~ 8.4\,(3)$ respectively.
  \label{fig:results}
\end{figure*}
We fit the set of tabulated \texttt{Cloudy} models to the observed and de-reddened line intensities from Table.~\ref{tab:lines}, after subtracting the line intensities expected for the shocked plasma around \src (also provided in Table.~\ref{tab:lines}).     

Independent feedback studies of the same system \citep{Abolmasov2008,Berghea2010,Berghea2012}, as well as our work here, demonstrate that the predicted line intensities depend strongly on the shape and integrated luminosity of the broad-band emission, and in particular the UV continuum. The data coverage in the UV, however, is still sparse. For our spectral description we assumed that the wind photosphere is quasi-spherical (Fig.~\ref{fig:sed}). We therefore have to consider this systematic uncertainty on top of the statistical uncertainties of the observed line intensities. Such systematics are hard to quantify. Numbers around tens of percent seem fair given the number of systematics and assumptions that enter our analysis, especially the choice of (spectral) models that can only be crude simplifications. After running the fits with 10\%, 15\%, and 30\% systematics for all five choices for $L_\mathrm{UV}$, we find that a value of 30\% allows us to reach statistically reasonable fits for at least one case of $L_\mathrm{UV}$. We consider fits to be reasonable if $\chi^{2}_\mathrm{red}$ is not larger than 2-3, if the $\chi^{2}$ space is not completely degenerate, and if constraints can be placed on the fit parameters $L_\mathrm{src,soft}$ and $L_\mathrm{src,hard}$. The results of the fits are shown in Table~\ref{tab:fitstat_cloudy} and in parts in Fig.~\ref{fig:results}.

\begin{table*}
  \centering
  \caption{Results of the fits using the series of \pycloudy\ table models assuming UV continua with varying luminosities. We list the shock-subtracted line intensities from Table~\ref{tab:lines}, and the modeled line intensities in units of $10^{-16}\,\mathrm{erg}\,\mathrm{cm}^{-2}\,\mathrm{s}^{-1}$, together with the data/model ratios in percentage units. The fit statistics are shown on the bottom row together with the best-fit input parameters, i.e., the implied intrinsic luminosities of the soft and hard X-ray components (\texttt{diskbb} and \texttt{nthcomp} respectively). All fit results in this table assumed an additional 30\% systematic uncertainty.}
  \label{tab:fitstat_cloudy}
  \begin{tabular}{llllllllllll} 
    \hline
     & & \multicolumn{2}{l}{$\mathrm{log}\,L_\mathrm{UV}=39.4$} & \multicolumn{2}{l}{$\mathrm{log}\,L_\mathrm{UV}=39.9$} & \multicolumn{2}{l}{$\mathrm{log}\,L_\mathrm{UV}=40.4$} & \multicolumn{2}{l}{$\mathrm{log}\,L_\mathrm{UV}=40.9$} & 
    \multicolumn{2}{l}{$\mathrm{log}\,L_\mathrm{UV}=41.4$} \\
    Line  &  $F_\mathrm{shock-subtracted}$  & $F_\mathrm{model}$ & ratio & $F_\mathrm{model}$ & ratio & $F_\mathrm{model}$ & ratio & $F_\mathrm{model}$ & ratio & $F_\mathrm{model}$ & ratio \\
      \hline
     \iontb{He}{ii}{4686} & $47\pm 2$ & 28 & 1.7 & 28 & 1.7 & 19 & 2.5 & 24 & 1.9 & 23 & 2.0\\
    \iontb{H}{$\beta$}{4861} & $90\pm4$ & 136 &  0.7 & 141 & 0.6 & 133 & 0.7 & 119 & 0.8 & 111 & 0.8 \\
    \iontbforb{O}{iii}{5007} & $1377\pm 30$ & 949 & 1.5 & 1223 & 1.1 & 1312 & 1.1 & 1674 & 0.8 & 1616 & 0.9 \\
    \iontbforb{N}{ii}{6583} & $417\pm12$ & 475 & 0.9 & 358 & 1.2 & 412 & 1.0 & 69 & 6.1 & 34 & 12.2 \\
    \iontbforb{S}{ii}{6716} & $249\pm3$ & 134 & 1.9 & 143 & 1.7 & 161 & 1.5 & 37 & 6.7 & 19 & 13.2 \\
    \iontbforb{S}{ii}{6731} & $163\pm4$ & 110 & 1.5 & 125 & 1.3 & 127 & 1.3 & 36 & 4.5 & 18 & 8.8 \\
    \hline
    \multicolumn{2}{l}{$\mathrm{log}\,L_\mathrm{X,soft}$} & \multicolumn{2}{l}{$40.7^{+1.2}_{-0.8}$} & \multicolumn{2}{l}{$40.1^{+1.5}_{-1.8}$} & \multicolumn{2}{l}{$39.9^{+1.9}_{-1.5}$} & \multicolumn{2}{l}{$38^{+4}_{-0}$} & \multicolumn{2}{l}{$41.5^{+0.9}_{-3.2}$} \\
    \multicolumn{2}{l}{$\mathrm{log}\,L_\mathrm{X,hard}$} & \multicolumn{2}{l}{$38^{+4}_{-0}$} &  \multicolumn{2}{l}{$38.8^{+3.2}_{-0.9}$} & \multicolumn{2}{l}{$41.4^{+0.6}_{-0.5}$} & \multicolumn{2}{l}{$42.0^{+0.0}_{-0.4}$} & \multicolumn{2}{l}{$41.98^{+0.00}_{-0.29}$}\\
        \multicolumn{2}{l}{Fit Statistics} & \multicolumn{2}{l}{$\chi^{2}\,(\mathrm{dof}) = 9.5\,(3)$} & \multicolumn{2}{l}{$\chi^{2}\,(\mathrm{dof}) = 8.2\,(3)$} & \multicolumn{2}{l}{$\chi^{2}\,(\mathrm{dof}) = 8.4\,(3)$} & \multicolumn{2}{l}{$\chi^{2}\,(\mathrm{dof}) = 26.5\,(3)$} & \multicolumn{2}{l}{$\chi^{2}\,(\mathrm{dof}) = 31.2\,(3)$} \\
    \hline
  \end{tabular}
\end{table*}

We can reproduce the MF16 nebula lines well, mostly within a deviation of a few tens of percent in line intensity, when considering an intrinsic UV continuum that has an integrated luminosity equal to or less than the inferred, observed value of $\mathrm{log}\,L_\mathrm{UV}=40.4$ (see Fig.~\ref{fig:results}). In general, the data/model ratios are largest for \heii, \siione, and \siitwo. The model nebulae simulated with \texttt{Cloudy} start to deviate significantly from the observed line intensities above a value of $\mathrm{log}\,L_\mathrm{UV}=40.4$ for the lines \nii, \siione, and \siitwo.   

\begin{figure*}
  \includegraphics[width=\textwidth]{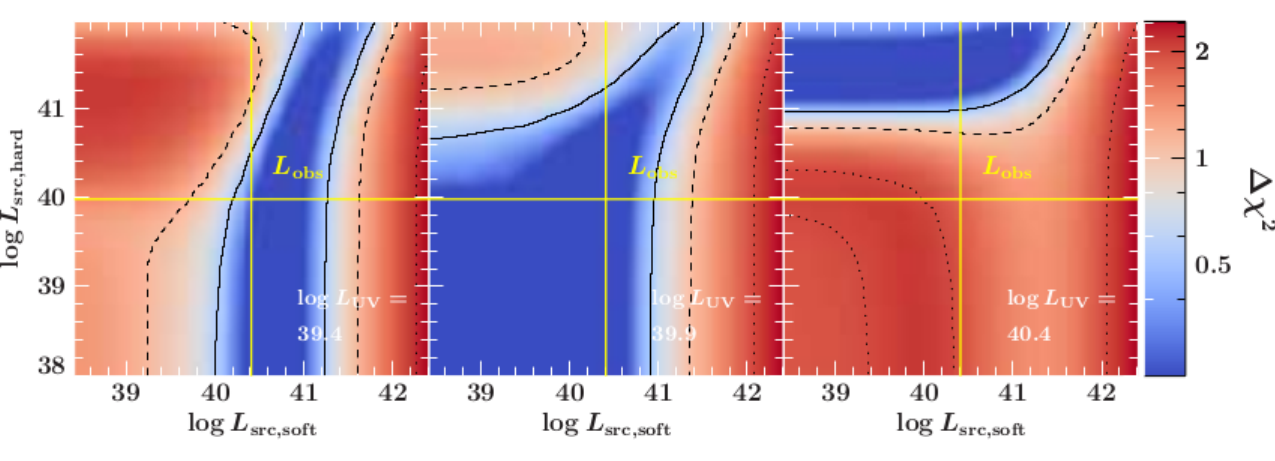}
  \caption{$\Delta \chi^2$ space for the three acceptable fits using our \cloudy-based table model. The fit parameters are the intrinsic source luminosities $L_\mathrm{src}$ for the soft and hard-X-ray model components of our continuum. We plot contours for the 68.3\%, 90\%, and 99\% confidence limits (dotted, dashed, and solid lines, respectively). The color scale ranges between $\Delta \chi^{2}=0$ and the maximum of each map for each panel, i.e., 2.3, 2.4, and 2.4, respectively. The beaming factor $\mathcal{B} = L_\mathrm{obs}/L_\mathrm{src}$ is larger than unity below the horizontal yellow line for the hard X-rays and to the left of the vertical yellow line for the soft X-rays. The intrinsic UV luminosity corresponds to $\mathrm{log}\,L_\mathrm{UV}=39.4, 39.9$ and $40.4$ from left to right. For all panels, we assume additional 30\% systematic uncertainties.}
  \label{fig:contour}
\end{figure*}

Table~\ref{tab:fitstat_cloudy} shows the constraints on the fitted parameters, i.e., the intrinsic luminosities of the soft and hard X-ray components \texttt{diskbb} and \texttt{nthcomp}, respectively. A key result of our analysis is that the fit is highly sensitive to changes in the luminosity of the UV spectrum that is irradiating the nebula whereas the component luminosities \lsrcsoft\ and \lsrchard\ are highly degenerate. This result becomes even more obvious when examining the largely degenerate $\chi^{2}$ parameter spaces for the fits with $\mathrm{log}\,L_\mathrm{UV}=39.4, 39.9, \mathrm{and}~40.4$ in Fig.~\ref{fig:contour}. All three cases for $L_\mathrm{UV}$ are statistically equally plausible.
While the soft X-ray luminosity seems to be well-constrained for the case of $\mathrm{log}\,L_\mathrm{UV}=39.4$, the hard X-ray spectral component is strongly degenerate. The opposite is true for $\mathrm{log}\,L_\mathrm{UV}=40.4$ which is close to the value we infer from our best guess for the de-reddened UV spectrum. The intermediate case of $\mathrm{log}\,L_\mathrm{UV}=39.9$ results in degenerate luminosities for both the soft and hard X-ray spectral components, and only upper limits can be inferred for $L_\mathrm{soft}$ and $L_\mathrm{hard}$.
A crucial result is that, especially at low UV luminosities, our fits to the observed nebula line intensities seem to be largely insensitive to changes in the hard-X-ray luminosity. 

\section{Discussion}
\label{sec:results} 
We have been able to reproduce the response of the MF16 nebula surrounding \src\, assuming intrinsic UV luminosities of $\mathrm{log}\,L_\mathrm{UV}=39.4, 39.9$ and $40.4$ (see Fig.~\ref{fig:results}). This range of UV luminosities is realistic given the uncertainties in the SED of \src\ (Fig.~\ref{fig:sed}). 

For the study, we assumed feedback both via photoionization and shock-ionization. While photoionization alone is able to reproduce the observed optical line intensities, several arguments indicate the presence of shocks: the multi-wavelength nebula morphology, the dynamics of MF16 measured via line widths, and the electron gas density of nearly 600\,cm$^{-3}$ -- which is much higher than the average ISM density of $1-10\,\mathrm{cm}^{-3}$ -- which argues for swept-up gas. Since the shock-ionization component can not be constrained statistically, we fix its contribution to the line intensities based on information on the shock speed and by using the \texttt{Mappings III} library.

We note that the spectral modeling of the X-ray continuum is somewhat ambiguous.
Due to the off-axis location of the X-ray source, we do not spatially resolve the nebula with \chandra\ and therefore cannot distinguish the correct model based on imaging data alone. However, the spectral-timing behavior described by \citet{Middleton2015a}, is consistent with X-rays being emitted from a super-critical disc with a wind, and the near-ubiquitous line-like residuals thought to imply the presence of winds \citet{Middleton2014, Middleton2015b} have been resolved into broadened atomic absorption and emission features \citep{Pinto2016,Pinto2017,Kosec2018,Kosec2021}, unambiguously associating ULXs with powerful outflows. As a result, we assume that the continuum irradiating the nebula originates from a super-critical disc \citep{Poutanen2007}. The soft spectrum of \src\ and its extreme UV brightness of $> 10^{39}\,$erg\,s$^{-1}$, is consistent with a ULX with an optically thick wind probably seen at fairly high inclinations.

With the above considerations in mind, we assess the impact of each component of the ULX's broad UV-to-X-ray SED on the nebula lines we observe. In particular, we wish to determine the possible emission pattern of the compact ULX. To do so, we test the response of the surrounding gas when changing the luminosities of the soft and hard-X-ray spectral components from Fig.~\ref{fig:sed}, i.e., thermal emission from the ULX wind, and emission from the inner parts of the accretion flow respectively. We then inspect the concordance between the observed luminosities $\log L_\mathrm{obs,soft}\sim 40.41$ and $\log L_\mathrm{obs,hard}\sim 39.98$ (yellow solid lines in Fig.~\ref{fig:contour}) and the statistically allowed intrinsic counterparts, i.e., \lsrcsoft\ and \lsrchard. We note that we have explicitly decoupled the UV and soft X-ray emission (assuming the former originates from the quasi-spherical photosphere at high inclinations).

We interpret our results in the framework of an optically thick wind with opening angle $\Omega$. While predominantly hard X-rays are prone to forward-scattering and therefore geometrical beaming in the line-of-sight \citep{King2009,Middleton2015a,Middleton2016}, soft-X-ray and UV photons should be emitted from larger solid angles (if not isotropically). The degree of geometrical beaming given by the so-called beaming factor $\mathcal{B}$ is typically defined by $\mathcal{B}=4\pi/\Omega \gtrsim 1$. However, by also defining $\mathcal{B} = L_\mathrm{obs}/L_\mathrm{src}$ we see that for face-on orientations, we would expect collimated emission to have $\mathcal{B} > 1$ but for increasing inclinations $\mathcal{B}$ would decrease and be less than 1 (typically for inclinations of $> 20^{\circ}-30^{\circ}$: \citealt{Dauser2017}). $\mathcal{B} < 1$ using this second definition (and where $L_\mathrm{src}$ can be inferred from our {\sc cloudy} modelling) must also imply anisotropy (and collimation for standard ULX models).   

Using the statistical constraints from Fig.~\ref{fig:contour}, for $\log L_\mathrm{UV}\sim 39.4$ and $39.9$, we find the beaming factor of the hard X-rays to be ambiguous, being able to take extreme values between $0.01 < \mathcal{B} < 100$ within the explored parameter space. For the soft X-ray component, we also find rather ambiguous beaming values with $0.2\lesssim \mathcal{B}\lesssim 2.6$ and $\mathcal{B}\gtrsim 0.3$ for $\log L_\mathrm{UV}\sim 39.4$ and $39.9$, respectively. For $\log L_\mathrm{UV}\sim 40.4$, the nebula seems sufficiently ionized and no additional soft X-ray contribution is needed (right panel of Fig.~\ref{fig:contour}). 

We can conclude from Fig.~\ref{fig:contour} that, for the cases of $\log L_\mathrm{UV}\sim 39.4$ and $39.9$, the hard emission could easily be beamed and have little-to-no impact on the optical lines, as the hard X-rays are much less capable of photoionizing the nebula than the soft X-rays. Only for the highest UV luminosity ($\log L_\mathrm{UV}\sim 40.4$), which is plausible based on this feedback study, do we find a very low beaming factor of $\mathcal{B}\lesssim 0.1$ for the hard X-ray component. In fact, most of the parameter space for $L_\mathrm{src, soft}$ and $L_\mathrm{src, hard}$ is forbidden in this case. This is not surprising -- the UV luminosity of $\log L_\mathrm{UV}\sim 40.4$ seems to significantly ionize the nebula without the need for extra ionizing soft X-ray photons. Only hard X-ray emission at luminosities in excess of that observed is statistically allowed and would need to be collimated and beamed away from our line-of-sight.

If the underlying assumption of an optically thick wind is correct, as supported by numerous observations of ULXs, numerical simulations and theoretical considerations would argue against significant beaming of soft X-rays \citep{Jiang2014,Poutanen2007,Middleton2015a,Narayan2017} but the hard X-rays would be expected to be more collimated. The wide allowed range of $\mathcal{B}_\mathrm{hard}$ from values of $\sim 0.3$ to arbitrarily high values would potentially support such an assertion. The same is true for $\mathcal{B}_\mathrm{soft}$, which we find to accept a wide range of values, while significant beaming also remains plausible given the constraints. The most important result of our feedback study is therefore less what it does constrain, but what it does not, i.e. nebula feedback cannot exclude strongly beamed, primarily hard X-ray emission. 

\section{Conclusions}

Using {\it Chandra} and {\it HST} data, we have obtained a time-averaged representation of the broad-band UV-to-X-ray spectrum of the ULX \src. To model this spectrum, we consider an optically thick wind resulting from super-Eddington accretion onto a stellar-mass compact object. Depending on the optical depth, photons can scatter off the wind to a lesser or greater extent and be geometrically collimated/beamed. In the standard model, harder X-ray photons will be scattered into a smaller solid angle than soft X-rays. To gain a better understanding of the geometry and emission characteristics of the accretion flow, we conducted a feedback study using the MF16 nebula as calorimeter. Using \cloudy\ and an irradiating broad-band SED, we obtained the photoionised response to compare to the observed optical line intensities (correcting for the nebular line response due to shock ionization).

We fitted a table model that returns optical line intensities given the intrinsic luminosities of the soft and hard spectral components as free parameters. This has allowed us to compare to the observed soft and hard luminosities and make inferences regarding the geometrical beaming of each of these components. We have found that -- as expected -- the response of the nebula line emission is strongly sensitive to changes in the irradiated UV flux. However the nebula responds far less to changes in the hard and soft X-rays and, from a statistical point of view, it is not possible to quantify the amount of beaming for either component. The major result of our study is that we {\it cannot} exclude strongly beamed X-ray emission based on nebula lines and it is not possible to argue for or against isotropic emission by only matching the integrated broad-band luminosity.

There are a number of assumptions within our analysis we bring the reader's attention to.  There are several sources of systematic uncertainty which challenge a precise quantification of the systematics: the radio and \halpha\ morphologies of the nebula imply shocked feedback to be prevalent, however, we can not self-consistently model both shock-ionization and photoionization. The exact intrinsic UV--hard-X-ray spectral shape is not well known, mainly due to strong systematics related to the de-reddening process; this eventually influences the nebular response. To address these points, we have assumed a 30\% systematic uncertainty on top of the statistical uncertainties of the observed optical line response of the nebula. Lastly, our approach makes the assumption that (a) the radiation emerges solely from a thick wind without any consideration for the presence of a neutron star with a strong dipole field, and (b) that the geometry of the wind can be approximated with a cone of opening angle $\Omega$. 

\section*{Acknowledgements}
TB is grateful for support from a NOVA (Netherlands Research School for Astronomy) Cross-Network grant. SM acknowledges support from an NWO (Dutch Research Council) VICI award, grant no. 639.043.513.  This work was also supported by the Australian government through the Australian Research Councils Discovery Projects funding scheme (DP200102471), and a grant from the Simons Foundation (00001470, PA). We made use of ISIS functions provided by ECAP/Remeis Observatory and MIT (\url{http://www.sternwarte.uni-erlangen.de/isis/}). We thank
J. E. Davis for the development of the \texttt{slxfig} module that has been used to prepare the figures in this work and P. Abolmasov for valuable comments. The National Radio Astronomy Observatory is a facility of the National Science Foundation operated under cooperative agreement by Associated Universities, Inc.

\section*{Data Availability}
The data underlying this article are either available in the article or can be shared on reasonable request to the corresponding author.



\bibliographystyle{mnras}

\nocite{*} 
\input{ulxfeedback_ngc6946_ulx1.bbl} 


\bsp	
\label{lastpage}
\end{document}